\documentclass[letterpaper]{JHEP3}

\usepackage{epsfig,amsmath,euscript}

\newcommand{\mcdot}{\!\cdot\!}

\newcommand{\vect}[1]{\mathbf{#1}}

\newcommand{\bra}[1]{\left\langle #1\right\rvert}
\newcommand{\ket}[1]{\left\lvert #1\right\rangle}

\def\X{\mathcal \chi}
\def\O{\mathcal O}
\def\l{\lambda}
\def\ll{\lambda^2}

\def\eps{\varepsilon}
\def\SCET{${\rm SCET}_{\rm /\!\!\!\!G}\,$}
\def\SCETG{${\rm SCET}_{\rm G}\,$}

\preprint{SI-HEP-2010-14}
\title{On Glauber modes in Soft-Collinear Effective Theory}
\author{Christian W. Bauer\\
Berkeley Center for Theoretical Physics, 
University of California, Berkeley, CA 94720, \\
Theoretical Physics Group, Lawrence Berkeley National Laboratory, 
Berkeley, CA 94720, \\
E-mail: \email{CWBauer@lbl.gov}}
\author{Bjorn O. Lange\\
Berkeley Center for Theoretical Physics, 
University of California, Berkeley, CA 94720, \\
Theoretical Physics Group, Lawrence Berkeley National Laboratory, 
Berkeley, CA 94720, \\
Theoretische Physik 1, Fachbereich Physik, Universit\"at Siegen, D-57068 Siegen, Germany,\\
E-mail: \email{BOLange@lbl.gov}}
 \author{Grigory Ovanesyan\\
Theoretical Division, T-2, MS B283, Los Alamos National Laboratory, Los Alamos, NM 87540,\\
E-mail: \email{ovanesyan@lanl.gov} }
 
\abstract{Gluon interactions involving spectator partons in collisions
  at hadronic machines are investigated. We find a class of examples
  in which a mode, called Glauber gluons, must be introduced to the
  effective theory for consistency. }

\begin{document}

\section{Introduction}

Factorization underlies any theoretical prediction at hadron
colliders, since it allows to separate the short distance,
perturbatively calculable physics from the non-perturbative
ingredients such as the parton distribution functions. First arguments in favor of factorization for hard QCD processes
to all order in perturbation theory appeared over three decades ago \cite{Amati:1978by,Libby:1978qf,Mueller:1978xu,Gupta:1979xj,Ellis:1978ty}.
Power counting, pinch analysis and physical pictures have been applied to analyze generic loop integrals at high energies in Refs.~\cite{Sterman:1978bi, Sterman:1978bj}, which led 
to the seminal work of Collins, Soper and Sterman on proofs of factorization theorems to all orders in perturbation series(see Refs.~%
\cite{Collins:1981ta,Collins:1981uk,Collins:1987pm,Collins:1989gx}
and references therein), and has been accomplished for many processes
of interest. In these proofs the Landau
equations~\cite{Landau:1959fi}, and their physical interpretation
using the Coleman-Norton Theorem~\cite{Coleman:1965xm}, were used to
identify the infrared singularities giving rise to the long distance
physics in arbitrary Feynman diagrams.

Singularities in Feynman diagrams arise if the integral over the loop
momenta leads to pinched singularities, where the contour of the loop
integration cannot be deformed to avoid the singularities in the
integrands. The Coleman-Norton Theorem allows to map these pinched
singularities onto configurations containing on-shell particles. For
most cases, pinched singularities are due to internal propagators
becoming ultrasoft or collinear to external particles, but there are
also pinched singularities associated with so-called ``Glauber" \cite{Bodwin:1981fv,Collins:1981ta}
modes\footnote{In Ref. \cite{Collins:1981ta} the same momentum region was called ``Coulomb".}. These are internal modes which have transverse momentum much in
excess of their longitudinal momentum, and can therefore not be
thought of as being on the mass-shell. In particular, Glauber gluons
connecting the remnants of the beams in hadronic collisions have been
argued to threaten factorization
\cite{Bodwin:1981fv,Collins:1981tt}. The presence of this transverse
interaction causes significant complication in the proof of the factorization for the Drell-Yan process. However, after summing over the unobserved hadronic final states, the effects from Glauber gluons cancel, and
factorization holds. This was explicitly shown to one loop order in \cite{Collins:1982wa} and generalized to higher orders in Refs. \cite{Bodwin:1984hc,Collins:1985ue}. Another approach was developed in Ref. \cite{Aybat:2008ct}, where the  cancelation of Glauber gluons in the inclusive cross section was shown
using the light-cone ordered perturbation theory.

Recently, a different approach to factorization proofs has emerged~\cite{Bauer:2002nz, Bauer:2002ie, Bauer:2003di, Manohar:2003vb,Gao:2005iu, Idilbi:2005er,Idilbi:2005ky, Idilbi:2005er, Chay:2005rz, Manohar:2005az, Becher:2006mr,Chen:2006vd, Lee:2006nr, Fleming:2007qr, Fleming:2007xt, Becher:2007ty,Schwartz:2007ib,Bauer:2008dt, Bauer:2008jx,Bauer:2010vu},
which is based on effective field theory techniques. Effective field
theories describe the long distance physics using a limited set of
degrees of freedom, with all short distance physics being integrated
out of the theory and contained in the Wilson coefficients of
operators. The relevant effective theory for collider physics is
soft-collinear effective theory
(SCET)~\cite{Bauer:2000ew,Bauer:2000yr,Bauer:2001ct,Bauer:2001yt} ,
which in its usual formulation contains collinear and ultrasoft degrees of
freedom. Factorization in SCET can be shown in a very straightforward
fashion, since one can show at the level of the Lagrangian that
the collinear and ultrasoft degrees of freedom decouple in SCET. The
advantage of the SCET approach to factorization theorems is that all
long-distance ingredients of the factorization theorem are defined as
matrix elements of effective theory operators, which allows to use renormalization
group equations to resum large logarithms that arise in most
perturbative expressions.

In its traditional formulation, SCET does not include Glauber
gluons. Given their importance in traditional factorization proofs it
is therefore crucial to understand if and how Glauber gluons enter the
effective theory framework.  An attempt to include Glauber gluons into
SCET was made in \cite{Liu:2008cc}, where the factorization of the DY
cross section in the presence of a Glauber mode was
reconsidered. However, there are several flaws in the arguments
presented by these authors, in particular their argument about the
necessity for Glauber gluons in SCET fails to consider the overlap
between ultrasoft, collinear and Glauber modes.  As we will show in
this work, a proper treatment of this overlap is crucial to understand
the contribution of Glauber gluons.  Note that the inclusion of
Glauber gluons into the SCET Lagrangian has also been used to describe
jet broadening in dense QCD matter \cite{Idilbi:2008vm,
  DEramo:2010ak}.

The main purpose of this paper is to investigate whether Glauber
gluons are required in SCET or not. We consider a well known operator
in SCET, the operator producing two energetic back-to-back quarks, and
perform the matching calculation determining its short distance Wilson
coefficient using two different choices for the external states. Given
that the short distance physics has to be independent of the external
states, a consistent effective theory has to give the same result for
both of the calculations. We will show that \SCET, the traditional
formulation without Glauber modes, gives different results for
different external states, while a theory that includes Glauber
gluons, \SCETG, will give the correct result for both choices. This
unambiguously shows that Glauber gluons need to be included for a
certain class of processes.  It should be pointed out, however, that
Glauber gluons may be integrated out of the effective theory, since
they can not be on their mass-shell and cannot appear as external
particles in perturbation theory, leading to a potential between pairs
of collinear fields in opposite directions \cite{StewartTalk}.

Having established the requirements of Glauber gluons in \SCETG, we
will study in some more detail the relationship between pinched
surfaces and the effective theory. Using a simple graphical
representation of the pinched surfaces one can easily understand the
necessity of Glauber gluons in the matching calculation under
consideration. We then proceed to use this analysis to derive what
final states are necessary to require Glauber modes to give a
non-zero contribution.

The plan of the paper is as follows. In Section \ref{sec:matching} we
perform a one-loop matching calculation within \SCET involving DY
amplitude topologies and show that the effective theory breaks down,
while \SCETG passes the consistency check. In Section
\ref{sec:Pinches} we perform the pinch analysis of a diagram with pure
spectator interactions and identify the right modes of an effective
theory for this process; then we discuss other processes where the
Glauber mode plays a role. Finally we conclude in Section
\ref{sec:concl}.

\section{An explicit matching calculation}
\label{sec:matching}
\begin{figure}[h!]
\begin{center}
\epsfig{file=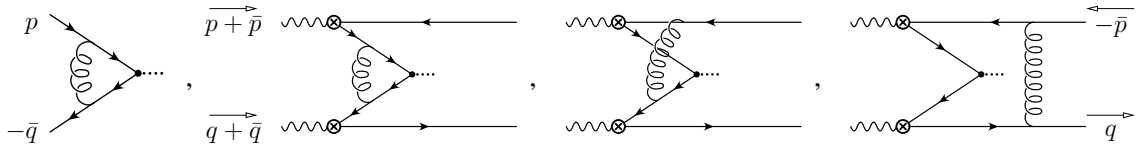, width=15cm}
\caption{\label{fig:figtopos} One-loop examples of $\langle \bar q
  q|\bar q \Gamma q|0\rangle$ (left) and $\langle \gamma^*
  \gamma^*|\bar q \Gamma q|\bar q q\rangle$ with active-active,
  spectator-active, and spectator-spectator interactions.}
\end{center}
\end{figure}

In this Section we will consider the well-studied SCET current $O_2 =
\bar \X_{\bar n} \Gamma \X_n$. The Wilson coefficient is usually
calculated by an explicit calculation using partonic external states
of free back-to-back quarks with large energy 
\begin{equation}
\label{standardmatching}
\langle \bar q q|\bar q \Gamma q|0\rangle = 
C_2 \langle \bar q q|O_2|0\rangle+\text{power corrections}\,.
\end{equation}
where $O_2$ is built of collinear gauge invariant fields
$\chi_n=W_n^{\dagger}\xi_n$. The Wilson line $W_n$ contains $n-$collinear gluons and $\xi_n$ 
is a two-component spinor describing an $n-$collinear quark field. (For more details see e.g.~Refs \cite{Bauer:2002nz,
  Bauer:2002ie, Bauer:2003di, Manohar:2003vb, Bauer:2008qu} and References therein).
The relevant Feynman diagram in the full theory is shown on the left
of Fig.~\ref{fig:figtopos}.

It is a well known consistency requirement of effective theories that the short distance
Wilson coefficient of any operators has to be independent of the long
distance physics in the process, and in particular it has to be
independent of the external states chosen for the matching
calculation. Inspired by the fact that Glauber gluons are known to
manifest themselves through interactions with beam remnants, we
intentionally choose more complicated external states, namely
\begin{equation} \label{newmatching}
\langle \gamma^* \gamma^*|\bar q \Gamma q|\bar q q\rangle = 
C_2 \langle \gamma^* \gamma^*|O_2| q \bar q  \rangle\,,
\end{equation}
where the two photons in the initial state and the two quarks in the 
final state are back-to-back with large energy. 
In this matrix element, each of the two off-shell photons converts
into a $q \bar q$ pair, with a quark and an antiquark from each photon
are then annihilated by the operator, while the other two quarks end
up in the final state. The required Feynman diagrams in the full
theory are depicted in Figure~\ref{fig:figtopos}. We refer to the
quarks that are annihilated by the operator as active, while the
quarks that end up in the final state are called spectators, and
divide the loop diagrams into active-active, active-spectator and
spectator-spectator diagrams, as indicated in the Figure.  Note that
beside the internal propagators that are not involved in the loop
integral, the active-active diagram is the same diagram as in the
matching calculation with a $q \bar q$ in the initial state, and
should by itself give the correct result for the matching
calculation. This implies that the sum of the contributions to $C_2$
from active-spectator and spectator-spectator diagrams has to yield
zero in a consistent effective theory.

For the purposes of this paper, i.e.~identifying contributing modes,
it is sufficient to study the singularity structure of the integrands,
and we will therefore omit the numerator structure of propagators for
simplicity. In the rest of this section we proceed as follows: The
Wilson coefficient $C_2$ is determined via a matching calculation of
the simple matrix element given in Eq.~(\ref{standardmatching}). Then we repeat the
procedure for the more complicated matrix element given in Eq.~(\ref{newmatching}).
First, we will calculate the contributions to the DY process in the
full theory straightforwardly. Next, we perform the matching
calculation by computing the corresponding diagrams in \SCET, finding
that $C_2$ is entirely reproduced by the active-active topology, as
expected. The spectator-active diagrams in \SCET match the
corresponding diagrams in the full theory. In the spectator-spectator topology, however, only the
soft diagram in \SCET is of leading power, and this diagram does {\em
  not} reproduce the full theory diagram. It appears that this
contribution would change both the Wilson coefficient and its
anomalous dimension from the findings of Eq.~(\ref{standardmatching}),
which is inconsistent.  Finally we consider \SCETG, which includes
Glauber gluons. Repeating the matching once again, we find that
Glauber gluons do not contribute for both Eq.~(\ref{standardmatching}) and
the active-active part of Eq.~(\ref{newmatching}), so that $C_2$ remains
unchanged. Interestingly, the active-spectator topologies in the full
theory and \SCETG again match identically due to the fact that the
naive contribution from Glauber gluons is precisely cancelled by its
overlap with the collinear gluon exchange. Finally the
spectator-spectator topologies also match, and \SCETG passes the
consistency check. We emphasize that the full theory diagram is
identically reproduced by the sum of the soft and the Glauber gluon
exchange, and not by the soft gluon exchange alone.

Before we consider the various diagrams in turn, let us set up some
notation. At tree level we define
\begin{equation}
\langle \bar q q|\bar q \Gamma q|0\rangle_{\rm tree} = 
\langle \bar q q|O_2|0\rangle_{\rm tree} = 1 \,.
\end{equation}
For the matching calculation denoted in Eq.~(\ref{standardmatching})
we denote the full theory matrix element at one-loop by
\begin{equation} \label{QCDTh1}
\langle \bar q q|\bar q \Gamma q|0\rangle_{\rm loop} = I_3 \,,
\end{equation}
where $I_3$ denotes the result of the 1-loop calculation. In
conventional \SCET, the gluon is either collinear or ultrasoft (which
we for brevity call soft everywhere below). They scale like
$(\ll,1,\l)$ and $(\ll,\ll,\ll)$ in their light-cone components,
respectively, where $\l$ is the small expansion parameter of SCET, and
the light-cone components are as usual with respect to the null
vectors $n$ and $\bar n$ in the $\pm z$ direction. To one loop order
the amplitude for this effective theory can therefore be written as
the sum of tree level and three effective theory diagrams
\begin{equation}
\langle \bar q q|O_2|0\rangle = 1 +  I_3^{c}+I_3^{\bar{c}}+I_3^{s} \,.
\end{equation}
Thus, the Wilson coefficient is given by
\begin{equation}
C_2 = \frac{\langle\bar q q|\bar q \Gamma q|0\rangle}{\langle\bar q q
|O_2|0\rangle} = 1+I_3 - (I_3^{c}+I_3^{\bar{c}}+I_3^{s}) \,.
\end{equation}

For the alternative external states, Eq.~(\ref{newmatching}), we assign
momenta $p+\bar p$ and $q+\bar q$ to the two initial virtual photons,
and $\bar p$ and $q$ to the two outgoing(spectator) partons. The
tree-level amplitude is simply given by the two propagators of the
active quarks, such that we find in both the full and the effective
theory
\begin{equation}
\bra{\gamma^*\gamma^*} \bar q \Gamma q\ket{\bar{q}q} _{\text{tree}}=  
\bra{\gamma^*\gamma^*} O_2\ket{\bar{q}q} _{\text{tree}}=
\frac{1}{p^2\bar q^2}\,.
\end{equation}
At one loop order, the full theory amplitude of the matrix element of
our interest is equal to sum of three topologies, which we write as
$I_3$, $I_4$, $I_5$, namely a triangle graph in the active-active, box
graphs in the spectator-active and a pentagon graph in the
spectator-spectator topology. We write for the full theory calculation
\begin{equation} \label{eq:ampdecomp}
  \bra{\gamma^*\gamma^*} \bar q \Gamma q \ket{\bar{q}q}=
\frac{1}{p^2\bar q^2} + \frac{1}{p^2\bar{q}^2} I_3+
\frac{1}{\bar{q}^2}I^{(n)}_4 +\frac{1}{p^2} I^{(\bar{n})}_4+I_5\,,
\end{equation}
where the prefactors of $1/p^2$, $1/\bar{q}^2$ take into account the
propagators that are independent of the loop-momenta. The $(n)$ and
$(\bar n)$ superscript on the $I_4$ denote if the active quark is in
the $n$ or $\bar n$ direction.  Note that the $I_3$ denotes exactly
the same integral as in Eq.~(\ref{QCDTh1}).

In conventional \SCET the one loop amplitude can be written as
\begin{equation}
   \bra{\gamma^*\gamma^*} O_2\ket{\bar{q}q}=\frac{1}{p^2\bar q^2} +
 \frac{1}{p^2\bar{q}^2 } (I_3^{c}+I_3^{\bar{c}}+I_3^{s})
+\frac{1}{\bar{q}^2}(I_4^{(n)c}+I_4^{(n)s})+\frac{1}{p^2} 
(I_4^{(\bar{n})\bar{c}}+I_4^{(\bar{n})s})+I^{s}_5\,,
\end{equation}
where the additional superscript $c$, $\bar c$ and $s$ on the
integrals denotes if the gluon is collinear in the $n$ direction,
collinear in the $\bar n$ direction or soft. Note that we have already
used the fact that some graphs, i.e.~$I_4^{(n)\bar c}$, 
$I_4^{(\bar n)c}$, $I_5^c$, $I_5^{\bar c}$, are power-suppressed. The
Wilson coefficient is therefore
\begin{eqnarray}
C_2 &=& 1+ I_3 - (I_3^{c}+I_3^{\bar{c}}+I_3^{s}) + p^2 \left[I^{(n)}_4 
- \left(I_4^{(n)c}+I_4^{(n)s}\right) \right] + \bar q^2 
\left[ I_4^{(\bar{n})} - \left(I_4^{(\bar{n})\bar{c}}+I_4^{(\bar{n})s}\right) 
\right] \nonumber\\
&& + p^2 \bar q^2 \left[ I_5 - I^{s}_5\right]\,.\label{c2EFT1}
\end{eqnarray}

In theory \SCETG, which adds Glauber gluons with momentum
scaling $(\ll,\ll,\l)$ to the conventional \SCET, there are new Feynman
diagrams present at one loop. Denoting the triangle, box and pentagon
diagrams with Glauber gluons in the loop by $I_3^{g}$, $I_4^{(n)g}$,
$I_4^{(\bar{n})g}$ and $I^{g}_5$, respectively, we can write
\begin{eqnarray}
 \bra{\gamma^*\gamma^*} O_2\ket{\bar{q}q}_G &=&  \frac{1}{p^2\bar q^2} + 
\frac{1}{p^2\bar{q}^2 } (I_3^{c'}+I_3^{\bar{c'}}+I_3^{s} + I_3^{g})+
\frac{1}{\bar{q}^2}(I_4^{(n)c'}+I_4^{(n)s}+I_4^{(n)g})\nonumber\\
 &&+\frac{1}{p^2} (I_4^{(\bar{n})\bar{c}'}+I_4^{(\bar{n})s}+ 
I_4^{(\bar{n})g})+I^{s}_5 + I^{g}_5 \,,
\end{eqnarray}
such that we can write
\begin{eqnarray}
C_2 &=& 1+ I_3 - (I_3^{c'}+I_3^{\bar{c}'}+I_3^{s}+ I_3^{g}) + p^2 \left[I^{(n)}_4 - 
\left(I_4^{(n)c'}+I_4^{(n)s}  + I_4^{(n)g}\right) \right] \nonumber\\
&& + \bar q^2 \left[ I_4^{(\bar{n})} - \left(I_4^{(\bar{n})\bar{c}'}+
I_4^{(\bar{n})s} +  I_4^{(\bar{n})g} \right) \right] + p^2 \bar q^2 
\left[ I_5 - \left( I^{s}_5 + I^{g}_5 \right)\right] \,.\label{c2EFT2}
\end{eqnarray}

Thus, the condition that the Wilson coefficient $C_2$ has to be
independent of the external states chosen allows us to determine if
the theory with or without Glauber gluons is correct. If 
\SCET were correct one would find
\begin{equation}
 p^2 \left[I^{(n)}_4 - \left(I_4^{(n)c}+I_4^{(n)s}\right) \right] + 
\bar q^2 \left[ I_4^{(\bar{n})} - \left(I_4^{(\bar{n})\bar{c}}+
I_4^{(\bar{n})s}\right) \right]+ p^2 \bar q^2 \left[ I_5 - I^{s}_5
\right]= 0\,,\label{consistencySCET}
\end{equation}
while the presence of Glauber gluons changes this condition to
\begin{eqnarray}
&&
 p^2 \left[I^{(n)}_4 - \left(I_4^{(n)c'}+I_4^{(n)s}  + I_4^{(n)g}\right) 
\right] + \bar q^2 \left[ I_4^{(\bar{n})} - \left(I_4^{(\bar{n})\bar{c}'}
+I_4^{(\bar{n})s} +  I_4^{(\bar{n})g} \right) \right] \nonumber\\
&&
\qquad \qquad
+ p^2 \bar q^2 
\left[ I_5 - \left( I^{s}_5 + I^{g}_5 \right)\right]= 0\,.
\label{consistencySCETG}
\end{eqnarray}
We will explicitly show below that \SCET does not
satisfy its consistency check Eq.~(\ref{consistencySCET}), while 
the effective theory which
contains Glauber gluons does satisfy Eq.~(\ref{consistencySCETG}). 
We will see that this difference between
\SCET and \SCETG happens only in the
spectator-spectator topology, i.e.~in the last terms of
Eq.~(\ref{consistencySCET}) and Eq.~(\ref{consistencySCETG}).

It is important to note that the collinear integrals differ between
the two different versions of SCET. It is well known that care has to
be taken when defining the contributions arising from collinear
gluons, since the various modes in the effective theory have overlap
regions which can lead to double counting.  When integrating over
collinear loop momenta in \SCET, part of the integration is over a
region in which the collinear momenta become soft. This would double
count the soft region and therefore has to be removed from the
collinear diagrams. This procedure is called ``zero-bin
subtraction''\cite{Manohar:2006nz}, and the collinear contributions in
Eqs.~(\ref{c2EFT1}) and (\ref{c2EFT2}) are all zero-bin subtracted. In
particular we have
\begin{equation}
I_k^{c}=\tilde{I}_k^{c}-(I_k^{c})_{0s}\,,\qquad I_k^{\bar{c}}=
\tilde{I}_k^{\bar{c}}-(I_k^{\bar{c}})_{0s}\,,
\end{equation}
where $\tilde{I}_k^{c}$ denotes the naive unsubtracted integrals.
When adding Glauber gluons, there are more overlapping regions,
requiring a more involved zero bin subtraction. We can write
\begin{eqnarray}
I_k^{c'} &=& \tilde{I}_k^{c}-\left[(I_k^{c'})_{0g}+(I_k^{c'})_{0s}-%
(I_k^{c'})_{0g0s}\right] \,,\nonumber\\
I_k^{\bar{c}'} &=& \tilde{I}_k^{\bar{c}}-\left[(I_k^{\bar{c}'})_{0g}+%
(I_k^{\bar{c}'})_{0s}-(I_k^{\bar{c}'})_{0g0s}\right] \,, \nonumber\\
I_k^{g}&=&\tilde{I}_k^{g}-(I_k^{g})_{0s}\,.\label{ZBeft22}
\end{eqnarray}

\subsection{Full Theory one loop calculation}

The active-active topology giving rise to $I_3$ in the full theory is
simply a standard scalar triangle integral. In $d=4-2\epsilon$
dimensions one finds\footnote{In all diagrams in this paper we assumed that off-shellness is positive: $p^2, \bar{p}^2, q^2, \bar{q}^2, (p+\bar{p})^2, (q+\bar{q})^2>0$.}
\begin{eqnarray}
I_3 &=& (-i) g^2\mu^{2\epsilon}  \int \frac{\text{d}^{d}%
l}{(2\pi)^{d}}\frac{1}{\left[l^2+i0\right]
\left[(l+p)^{2}+i0\right]\left[(l-\bar{q})^{2}+i0\right]} \\
&=& \frac{\alpha_s}{4\pi}\, \frac{1}{p^+\bar{q}^-}
\left(\frac{\pi^2}{3}+\ln\frac{p^2}{p^+\bar{q}^- }
\ln\frac{\bar{q}^2}{p^+\bar{q}^-}\right)+{\cal{O}}
\left(\epsilon,\, {\lambda^2}\right).
\end{eqnarray}
The spectator-active topology can be written as the box integral
\begin{eqnarray}
I_4^{(n)} &=& (-i) g^2\mu^{2\epsilon}  \int \frac{\text{d}^{d}l}{(2\pi)^{d}}%
\frac{1}{\left[l^2+i0\right] \left[(l-\bar{p})^{2}+i0\right]
\left[(l+p)^{2}+i0\right] \left[(l-\bar{q})^{2}+i0\right]} \\
&=& \frac{\alpha_s}{4\pi}\, \frac{1}{\bar{q}^{-}}\,
\frac{1}{\bar{p}^2 p^{+} +p^2 \bar{p}^{+}} 
\left\{\frac{\pi^2}{3}-2\,\text{Li}_2\left(
{-\frac{p^2 \bar{p}^+}{\bar{p}^2{p}^+}}\right) \right. \nonumber\\
&& \qquad  \left. +\left[\ln
\left(\frac{\bar{p}^2 p^{+}}{p^2 \bar{p}^+}\right)-i\,\pi\right]\, 
\ln\left(\frac{\bar{q}^{-}\left(p^+\bar{p}^2+\bar{p}^{+}p^2\right)^2}%
{\bar{q}^2(p+\bar{p})^2p^+ \bar{p^2}}\right) \right\}+{\cal{O}}
\left(\epsilon, \lambda^{0}\right).
\end{eqnarray}
An analogous expression is valid for the second spectator-active
integral $I_4^{(\bar{n})}$.

The spectator-spectator topology giving rise to $I_5$ in the full
theory can be calculated via a pentagon integral which by standard 
procedures can be reduced to sum of five box integrals. The result is\footnote{This pentagon loop integral and also integrals $I_5^g, I_4^{(n)g}, (I_4^{(n)c'})_{0 g}$ have been calculated for a simplified case of  ``$\perp$"-less kinematics: $\vect{p}_{\perp}=\vect{\bar{p}}_{\perp}=\vect{{q}}_{\perp}=\vect{\bar{q}}_{\perp}=\vect{0}$. Since we are free to choose any external states in the matching calculation we can easily achieve these conditions.}
\begin{eqnarray}
I_5 &=& (-i) g^2\mu^{2\epsilon}  \int \frac{\text{d}^{d} l}{(2\pi)^{d}}
\frac{1}{\left[l^2+i0\right] \left[(l-\bar{p})^2+i0\right]
\left[(l+p)^2+i0\right] \left[(l-\bar{q})^2+i0\right]
\left[(l+q)^2+i0\right]} \nonumber\\
&=&\frac{\alpha_s}{4\pi} M^+ M^-
\Bigg\{\frac{1}{p^{+}\bar{p}^{+} (p+\bar{p})^2q^{-}\bar{q}^{-}
(q+\bar{q})^2} \Bigg[\ln \frac{\bar{p}^{+}p^2}{p^{+}\bar{p}^2} \, 
\ln\frac{q^{-}\bar{q}^2}{\bar{q}^{-}q^2}
\nonumber\\
&&
+i\pi\ln\frac{\bar{p}^2 
p^2 \bar{q}^2 q^2}{\bar{p}^{+}p^{+}\bar{q}^{-}q^{-}(M^+M^-)^2}
+\pi^2\Bigg] 
+\frac{2\pi i \,}{p^{+}\bar{p}^{+}(p+\bar{p})^2(M^-)^2-q^{-}
\bar{q}^{-}(q+\bar{q})^2(M^+)^2}
\nonumber\\
&& 
\times \Bigg[\frac{(M^-)^2\ln\frac{M^+ (M^-)^3}{q^{-}\bar{q}^{-}
(q+\bar{q})^2}}{q^{-}\bar{q}^{-}(q+\bar{q})^2}
-\frac{(M^+)^2\ln\frac{(M^+)^3 M^-}{p^{+}\bar{p}^{+}(p+
\bar{p})^2}}{p^{+}\bar{p}^{+}(p+\bar{p})^2}\Bigg]\Bigg\}
+ {\cal{O}}\left(\epsilon,\, \frac{1}{\lambda^2}\right)
\,,
\end{eqnarray}
where we have defined 
\begin{eqnarray}
   M^+=p^+ + \bar{p}^+, \,\,\,\quad M^-=q^- + \bar{q}^- \,.
\end{eqnarray}

To determine the Wilson coefficient $C_2$ we need to subtract the
effective field theory diagrams from this result, which should 
cancel all IR divergences and reproduce the scalar one-loop result
\begin{equation}
C_2 = 1 + \frac{\alpha_s}{4\pi}
\Bigg[\frac{1}{\epsilon^2}+\frac{\ln\frac{\mu^2}{p^+\bar{q}^-}+i\pi}{\epsilon}+\frac{1}{2}\ln^2\frac{\mu^2}{p^+\bar{q}^-}+i\pi\ln\frac{\mu^2}{p^+\bar{q}^-}-\frac{7}{12}\pi^2\Bigg]\,.
\end{equation}

\subsection{Conventional SCET: soft and collinear gluon exchanges}

The loop integrals for the effective theory modes in each topology can
be found trivially by expanding the integrands in the full theory with
the appropriate scaling of the gluon momenta before integrating over
the loop momentum. 
The zero-bin subtraction integrals can be found similarly by
expanding the effective theory loop integrals with the scaling of the
overlap mode. All zero-bin integrals are scaleless in this effective
theory, so they just convert infrared poles into the ultraviolet ones.

The results for all diagrams in \SCET are given by
\begin{eqnarray}
I_3^{c} &=&
\frac{\alpha_s}{4\pi}\,\frac{1}{p^+\bar{q}^-}\left[
-\frac{1}{\epsilon^2}
-\frac{\ln\frac{\mu^2}{p^2}+i\pi}{\epsilon}
-\frac{1}{2}\left(\ln\frac{\mu^2}{p^2}+i\pi\right)^2
+\frac{\pi^2}{12}\right]\,,
\nonumber\\
I_3^{\bar{c}} &=&
\frac{\alpha_s}{4\pi}\,\frac{1}{p^+\bar{q}^-}\left[
-\frac{1}{\epsilon^2}
-\frac{\ln\frac{\mu^2}{\bar{q}^2}+i\pi}{\epsilon}
-\frac{1}{2}\left(\ln\frac{\mu^2}{\bar{q}^2}+i\pi\right)^2
+\frac{\pi^2}{12}\right]\,,
\nonumber\\
I_3^{s} &=&
\frac{\alpha_s}{4\pi}\,\frac{1}{p^+\bar{q}^-}\left[
\frac{1}{\epsilon^2}
+\frac{1}{\epsilon} \ln\frac{\mu^2 p^+\bar{q}^-}{p^2\bar{q}^2}+\frac{i\,\pi}{\epsilon}
+\frac{1}{2}\ln^2\frac{\mu^2 p^+\bar{q}^-}{p^2\bar{q}^2}
+i\,\pi\ln\frac{\mu^2
p^+\bar{q}^-}{p^2\bar{q}^2}
-\frac{\pi^2}{4} \right]\,,
\nonumber\\
I_4^{(n)c} &=&
\frac{\alpha_s}{4\pi}\,\frac{1}{\bar{q}^{-}}\mcdot\frac{1}{\bar{p}^2 p^{+} +p^2 \bar{p}^{+}}\Bigg\{
\frac{\ln\frac{p^2 \bar{p}^+}{\bar{p}^2 p^+}+i\pi}{\epsilon} 
-\frac{7\pi^2}{6}
-2\,\text{Li}_2{\left(-\frac{p^2 \bar{p}^+}{\bar{p}^2{p}^+}\right)}
\nonumber\\
 &&
+i\pi \ln\frac{\mu^2 p^2(p+\bar{p})^2 (\bar{p}^+)^2}{(\bar{p}^2 p^{+} +p^2 \bar{p}^{+})^2 \bar{p}^2}
+ \ln\frac{\bar{p}^2 p^+}{p^2 \bar{p}^+ }\left[
	\ln\frac{(\bar{p}^2 p^{+} +p^2 \bar{p}^{+})^2}{ (p+\bar{p})^2 p^+ \bar{p}^+\bar{p}^2}
	-\frac{1}{2}\ln\frac{\mu^4 p^+}{p^2\bar{p}^2\bar{p}^+}\right]\Bigg\}\,,
\nonumber\\
I_4^{(n)s} &=&
\frac{\alpha_s}{4\pi}\,\frac{1}{\bar{q}^{-}}\mcdot\frac{1}{\bar{p}^2 p^{+} +p^2 \bar{p}^{+}}\Bigg[ 
-\frac{\ln\frac{p^2 \bar{p}^+}{\bar{p}^2 p^+}+i\pi}{\epsilon}
+\frac{1}{2}\ln\frac{\bar{p}^2 p^+}{p^2\bar{p}^+}\ln\frac{\mu^4 p^+\bar{p}^+(\bar{q}^-)^2}{p^2\bar{p}^2(\bar{q}^2)^2}
\nonumber\\
&& \hspace{4cm}
-i\pi\ln\frac{\mu^2 p^2 (\bar{p}^+)^2\bar{q}^-}{\bar{q}^2(\bar{p}^2)^2
p^+}+\frac{3}{2}\pi^2\Bigg]\nonumber,\\
I_5^{s} &=&
\frac{\alpha_s}{4\pi} \frac{M^+ M^-}{p^{+}\bar{p}^{+} (p+\bar{p})^2 q^{-}\bar{q}^{-} (q+\bar{q})^2}\Bigg[
-\frac{2 \,i\pi}{\epsilon}
+\ln\frac{\bar{p}^{+}p^2}{p^{+}\bar{p}^2}\ln\left(\frac{q^{-}\bar{q}^2}{\bar{q}^{-}q^2}\right)
\nonumber\\
&&
\hspace{4cm}
+i\pi\ln\left(\frac{\bar{p}^2
p^2 \bar{q}^2
q^2}{\bar{p}^{+}p^{+}\bar{q}^{-}q^{-}\mu^4}\right)+3\pi^2\Bigg]\,.
\end{eqnarray}

We can combine the results from the full theory and \SCET calculation to find
\begin{eqnarray}
I_3-I_3^{s}-I_3^{c}-I_3^{\bar{c}} &=&
\frac{\alpha_s}{4\pi}
\Bigg[\frac{1}{\epsilon^2}+\frac{\ln\frac{\mu^2}{p^+\bar{q}^-}+i\pi}{\epsilon}+\frac{1}{2}\ln^2\frac{\mu^2}{p^+\bar{q}^-}+i\pi\ln\frac{\mu^2}{p^+\bar{q}^-}-\frac{7}{12}\pi^2\Bigg]\,,
\nonumber\\
p^2 \left({I_4-I_4^{(n)s}-I_4^{(n)c}}\right)&=& 0 \,,
\nonumber\\
p^{2}\bar{q}^2\left({I_5-I_5^{s}}\right)  &=& 
\frac{\alpha_s M^+ M^-}{4\pi}\Bigg\{
\frac{p^{2}\bar{q}^2}{p^{+}\bar{p}^{+}(p+\bar{p})^2q^-\bar{q}^-(q+\bar{q})^2}
\left[
\frac{2\pi i}{\epsilon}
-2\pi^2
+2\pi \,i \,\ln\frac{\mu^2}{M^+ M^-}\right]
\nonumber\\
&&
+\frac{2\pi i \,p^{2}\bar{q}^2}
{p^{+}\bar{p}^{+}(p+\bar{p})^2 (M^{-})^2-q^{-}\bar{q}^{-}(q+\bar{q})^2 (M^+)^2}
\nonumber\\
&& \qquad
\times 
\left[(M^-)^2\frac{\ln\frac{(M^-)^3 M^+}{q^{-}\bar{q}^{-}(q+\bar{q})^2}}{q^{-}\bar{q}^{-}(q+\bar{q})^2}
-(M^+)^2\frac{\ln\frac{(M^+)^3 M^-}{p^{+}\bar{p}^{+}(p+\bar{p})^2}}{p^{+}\bar{p}^{+}(p+\bar{p})^2}\right]\Bigg\}\,.\,
\end{eqnarray}
As discussed before, the triangle diagrams are the same as with the
$q\bar q$ external state, such that the combination
$I_3-I_3^{s}-I_3^{c}-I_3^{\bar{c}}$ reproduces the known Wilson
coefficient. While the box diagrams are the same in the full and
effective theory and therefore the combination
${I_4-I_4^{(n)s}-I_4^{(n)c}}$ is equal to zero, this is not true for
the combination $I_5-I_5^{s}$. This implies that the Wilson
coefficient $C_2$ is different when calculated with the two off-shell
photons in the initial state, which shows clearly that \SCET does not
reproduce the infrared physics of the full theory. On top of that, we
can see that the combination $I_5-I_5^{s}$ is UV divergent as well,
indicating another failure of \SCET.

\subsection{SCET including Glauber gluons}

Having found a serious problem in \SCET we now repeat our
matching calculation including Glauber gluons. One interesting aspect
of this theory is there exists a zero-bin contribution in the spectator-active
topology that does not vanish, such that it is more than merely a ``pull-up''
contribution \cite{Manohar:2006nz}. The non-vanishing overlap is
\begin{equation}
(I_4^{(n)c'})_{0 g} =
\frac{\alpha_s}{4\pi}\frac{1}{\bar{q}^-}\frac{1}{\bar{p}^+p^2+p^+
\bar{p}^2}\left[\frac{i\pi}{\epsilon}-i\pi
\ln\left(-\frac{\bar{p}^+p^2+p^+
\bar{p}^2}{\mu^2\left(p^++\bar{p}^+\right)}-i0\right)\right].
\end{equation}

Continuing with the calculation of the other contributions, we find
that several of the integrals are either equivalent to the \SCET case
or zero
\begin{equation}
I_3^{c'}=I_3^{c}\,, \qquad I_3^{\bar{c}'} =I_3^{\bar{c}}\,, \qquad I_3^{g}=0\,.
\end{equation}
For the remaining integrals we find
\begin{eqnarray}
I_4^{(n)c'} &=& I_4^{(n)c}-(I_4^{(n)c'})_{0 g}\,,\nonumber\\
I_4^{(n)g} &=&(I_4^{(n)c'})_{0 g}\,,\nonumber\\
I_5^{g} &=& 
\frac{\alpha_s}{4\pi}\Bigg\{
\frac{M^+ M^-}{p^{+}\bar{p}^{+}(p+\bar{p})^2q^-\bar{q}^-(q+\bar{q})^2}\left[
\frac{2\pi i}{\epsilon}
-2\pi^2
+2\pi \,i \,\ln\frac{\mu^2}{M^+ M^-}\right]
\nonumber\\
&&\qquad\qquad
+\frac{2\pi i \,M^+ M^-}
{p^{+}\bar{p}^{+}(p+\bar{p})^2 (M^{-})^2-q^{-}\bar{q}^{-}(q+\bar{q})^2 (M^+)^2}
\nonumber\\
&& \qquad\qquad\qquad
\times 
\left[(M^-)^2\frac{\ln\frac{(M^-)^3 M^+}{q^{-}\bar{q}^{-}(q+\bar{q})^2}}{q^{-}\bar{q}^{-}(q+\bar{q})^2}
-(M^+)^2\frac{\ln\frac{(M^+)^3 M^-}{p^{+}\bar{p}^{+}(p+\bar{p})^2}}{p^{+}\bar{p}^{+}(p+\bar{p})^2}\right]\Bigg\}\,.
\end{eqnarray}

The corresponding contributions to the Wilson coefficient $C_2$ from
different topologies are equal to
\begin{eqnarray}
I_3-I_3^{c'}-I_3^{\bar{c}'}-I_3^{g}-I_3^{s} &=&
\frac{\alpha_s}{4\pi}
\Bigg[\frac{1}{\epsilon^2}+\frac{\ln\frac{\mu^2}{p^+\bar{q}^-}+i\pi}{\epsilon}+\frac{1}{2}\ln^2\frac{\mu^2}{p^+\bar{q}^-}+i\pi\ln\frac{\mu^2}{p^+\bar{q}^-}-\frac{7}{12}\pi^2\Bigg]\,,
\nonumber\\
p^2\left({I_4-I_4^{(n)c'}-I_4^{(n)g}-I_4^{(n)s}}\right) &=& 0,
\nonumber\\
p^2\bar{q}^2\left({I_5-I_5^{s}-I_5^{g}}\right) &=& 0\,.
\end{eqnarray}
Thus, \SCETG with the inclusion of Glauber gluons does reproduce the
correct Wilson coefficient, which can be taken as a strong indication
that it is the correct effective theory.

\section{Pinch analysis and power counting}\label{sec:Pinches}

\subsection{The spectator-spectator interaction in Drell-Yan}

\begin{figure}[t]
\begin{center}
\epsfig{file=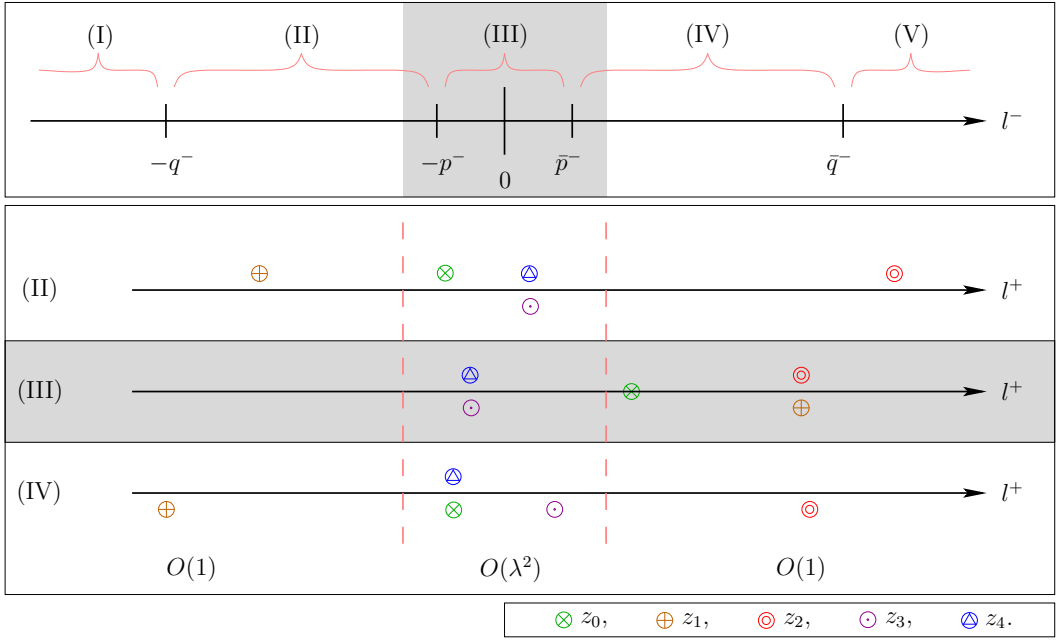, width=14cm}
\caption{\label{fig:axis}A few representative ``constellations'' of
  singularities during the integration over $l_\perp$. We show the
  three non-trivial intervals in $l^-$ (regions (II), (III), (IV)
  above) and one sample projection onto the complex $l^+$ plane for
  each region. The poles in the complex $l^+$ plane can be pinched in
  different locations, depending on the values of $l^-$ and $l_\perp$.
  For example, in region (III) $l^-$ counts as $O(\ll)$, and there are
  two pinched singularities: first, there exists a value of $l_\perp$
  for which $l^+$ is pinched between $z_3$ and $z_4\sim O(\ll)$;
  second, there exists a (generally different) value of $l_\perp$ for
  which $l^+$ is pinched between $z_1$ and $z_2\sim O(1)$. }
\end{center}
\end{figure}

As we showed in the previous Section, the spectator-spectator
contribution required a Glauber gluon for the effective theory to
reproduce the full theory result. In this Section we will investigate
the structure of this contribution in more detail in order to get more
insight into the required modes. The scalar integral contributing to
the spectator-spectator diagram is given by
\begin{equation}\label{eq:SSdiag}
I_5 = (-i)\int \frac{d^4 l}{(2\pi)^4} 
\frac{1}{l^2+i 0}
\frac{1}{(l+p)^2+i0}
\frac{1}{(l-\bar p)^2+i0}
\frac{1}{(l+q)^2+i0}
\frac{1}{(l-\bar q)^2+i0}\,.
\end{equation}
Decomposing the loop momentum $l$ into its light-cone components we
arrive at the following form, which is suitable for first integrating
over the $+$ component by contours and leaving the $\perp$ components
as a final integration:
\begin{equation}
\label{eq:Dfull}
I_5 =  (-i)\frac12 \int \frac{d^2 l_\perp}{(2\pi)^2} \int \frac{d l^-}{2\pi} N^-(l^-) \int \frac{dl^+}{2\pi}
\prod_{i=0}^4 \frac{1}{l^+ - z_i(l^-,l_\perp)}\,,
\end{equation}
where $\left[N^-(l^-)\right]^{-1} = l^-(l^-+p^-)(l^--\bar
p^-)(l^-+q^-)(l^--\bar q^-)$.  The positions of the singularities
$z_i$ in the complex $l^+$ plane are functions of $l^-$ and $l_\perp$,
as well as the external momentum components. Explicitly they are given
by
\begin{eqnarray}
z_0(l^-,l_\perp) &=& \frac{l_\perp^2-i 0}{l^-} \,, \nonumber\\
z_1(l^-,l_\perp) = \frac{(l_\perp+p_\perp)^2-i0}{l^-+p^-}-p^+\,, &\quad&
z_3(l^-,l_\perp) = \frac{(l_\perp+q_\perp)^2-i0}{l^-+q^-}-q^+\,, \nonumber\\
z_2(l^-,l_\perp) = \frac{(l_\perp-\bar p_\perp)^2-i0}{l^--\bar p^-}+\bar p^+\,, &\quad&
z_4(l^-,l_\perp) = \frac{(l_\perp-\bar q_\perp)^2-i0}{l^--\bar q^-}+\bar q^+\,.
\end{eqnarray}
Note that the locations of the poles above or below the real axis
changes during the integration over $l^-$ at the transitions $l^- =
-q^-\sim \O(1)$, $\bar q^- \sim \O(1)$, $-p^- \sim \O(\ll)$, $\bar
p^-\sim \O(\ll)$ and $l^- = 0$. For $l^- < -q^-$ or $l^- > \bar q^-$,
all poles are either above or below the real axis, such that the total
integral vanishes. We can divide the remaining range of $l^-$ into
three regions $-q^- < l^- < -p^-$, $-p^- < l^- < \bar p^-$ and $p^- <
l^- < \bar q^-$, as depicted schematically in Figure~\ref{fig:axis}.

We begin by considering the region $-p^- < l^- < \bar p^-$, labeled
(III) in the Figure. In this case the poles in $z_1$ and $z_3$ are
below the real axis, while the poles in $z_2$ and $z_4$ are above the
real axis. The pole in $z_0$ is either above or below the real axis%
\footnote{In the Figure we chose to indicate this by placing the pole
  directly on the real axis.}, depending on the sign of $l^-$. 

\begin{figure}[t!]
\begin{center}
\epsfig{file=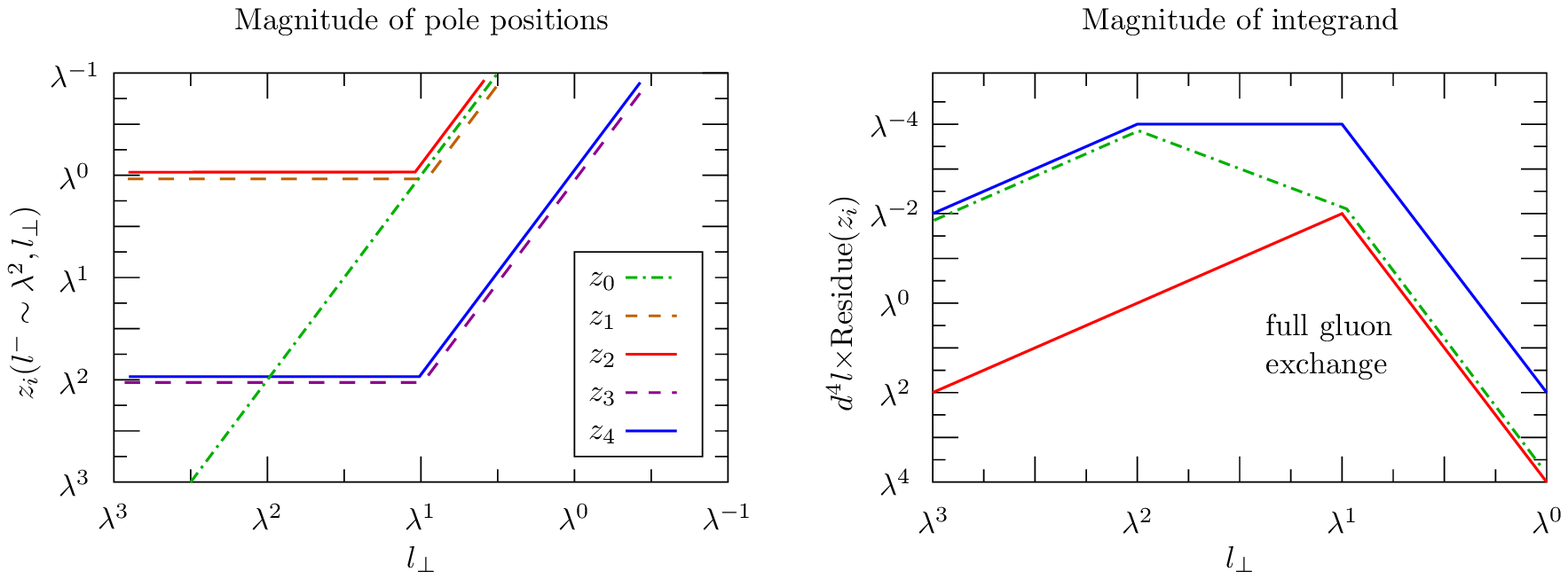, width=\textwidth} \\[2mm]
\epsfig{file=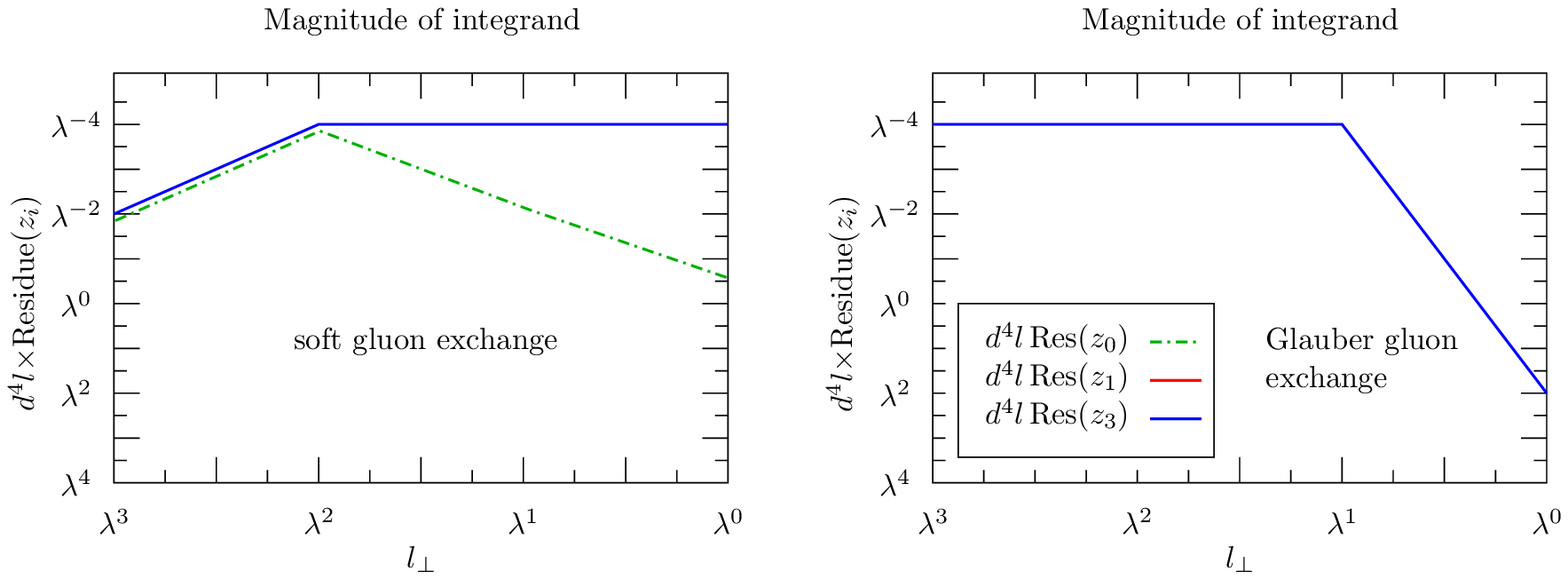, width=\textwidth}
\caption{\label{fig:poles}Top Left: magnitude of pole locations as a
  function of $l_\perp$. Dashed lines denote poles in the lower half
  plane, while solid ones are in the upper half plane. The dash-dotted
  line may be on either side. Top Right: magnitude of the residues of
  poles in the upper half plane (including integration measures). The
  color coding is identical to the one on the left. Lower Row:
  magnitude of the residues for the effective modes soft (left) and
  Glauber (right).}
\end{center}
\end{figure}

Since $l^- \sim O(\ll)$ the pole location $z_i$, and hence the
magnitude of $l^+$ after its residue is taken, depends only on the size of
$l_\perp$. In particular, $z_0 \sim O(l_\perp^2 / \ll)$ depends
strongly on the power assignment of $l_\perp$, which demonstrates the
shortcomings of the cartoon in Figure~\ref{fig:axis}. We find that a
new approach to illustrate the $l_\perp$ dependence is in order, which
we present in Fig.~\ref{fig:poles}.

The top left part of Fig.~\ref{fig:poles} shows the magnitude of the
pole position $z_i$ as a function of $l_\perp$ on a double logarithmic
scale. The green, orange, red, magenta and blue curves correspond to
$z_0$, $z_1$, $z_2$, $z_3$ and $z_4$, respectively. We also visualize
the sign of the pole by showing the $z_i$ above the
real axis by solid lines, those with poles below the axis by dashed
lines, and $z_0$, whose pole that can change location, by the
dot-dashed line. Whenever a solid and a dashed line are present at the
same point, the two corresponding poles can be pinched. For example, for
$l_\perp \sim O(\l)$ there is a pinch between $z_3$ and $z_4$ leading
to $l^+ \sim \lambda^2$, but also a pinch between $z_0$, $z_1$ and
$z_2$ leading to $l^+ \sim 1$.

Next, for a given $l_\perp$, we can also determine the magnitude of
the contribution of each of the residues to the integrand of $I_5$. In
the top right part of Fig.~\ref{fig:poles} we illustrate the magnitude
of the integral, taking into account the scaling of the $d^4 l$. In
this Figure we have chosen to close the contour above the real axis,
so that the poles below the real axis do not contribute. Note that
leading power corresponds to $\lambda^{-4}$, consistent with
Eq.~(\ref{eq:ampdecomp}).

These diagrams can be used to identify the modes required in an
effective theory. From the top right plot we see that a leading order
contribution $\lambda^{-4}$ can only come from the residue of $z_4$,
and that $l_\perp$ has to be between $\lambda$ and $\lambda^2$. The
residue from $z_2$ always leads to a power suppressed contribution, in
agreement with the explicit calculation performed in the previous
Section. From the top left plot we see that the corresponding poles
are indeed pinched and we also read off that for these cases the
momentum component $l^+$ scales as $\lambda^2$. Thus, the two scalings
of the loop momentum that give rise to leading order effects in the
spectator-spectator contribution are
\begin{equation}
l^\mu \sim (\lambda^2, \lambda^2, \lambda^2)\,, \qquad {\rm and} \qquad 
l^\mu \sim (\lambda^2, \lambda^2, \lambda)\,,
\end{equation}
which are the soft and Glauber mode. We will come back to the
possibility of $l^\perp$ being somewhere in between $\lambda$ and
$\lambda^2$ later.  

So far we have only considered the case $l^- \sim O(\ll)$, and in
particular the region $l^- \in [-p^-,\bar p^- ]$. Outside that
region one of the poles $z_1, z_2$ will cross the real axis onto the
other half plane, unpinching a singularity, which only results in the
removal of the already power-suppressed (collinear) contribution of its
residue. For completeness we shall also mention that the case $l^-
\sim O(1)$ will again lead to no further leading-power
contributions.
 
It is an easy exercise to draw plots analogous to the top right plot of
Figure~\ref{fig:poles} for the effective theory. As one obtains the
integrals in the effective theory by expanding the full integral about
the loop momentum modes, one similarly finds that the top right plot
in Figure~\ref{fig:poles} is simply expanded around the modes, which
is shown in the lower row of the Figure. In this case, expanding the
integrand around soft loop momentum will reproduce the plot below
$l_\perp\sim O(\ll)$ and continue the straight segment to the right of
it indefinitely. This is shown in the lower left plot.  Thus, the soft
contribution reproduces the infrared region and picks up a
$1/\eps_{\rm UV}$ singularity. On the other hand, an expansion around
the Glauber scaling (lower right plot) will reproduce the region
around $l_\perp \sim \l$ and pick up a $1/\eps_{\rm IR}$ pole. The
overlap between those two modes is obtained by expanding the Glauber
integral around the soft limit, which is a scaleless integral and
provides the subtraction of both poles, thus reproducing the full
theory at leading power. 

In the above we have only considered the two discrete choices $l_\perp =
\lambda$ or $l_\perp = \lambda^2$. However, the Figures seem to
indicate that any scaling of $l_\perp$ between these two extremes
leads to a leading order contribution and has a pinched pole. The
question then arises why we do not have to consider more modes, such
as one with momentum scaling as $l^\mu \sim (\lambda^2, \lambda^2,
\lambda^{3/2})$. Expanding around this point, we would obtain a mode
in the effective theory similar to the construction discussed
above. The important point, however, is that the overlap contributions
between this mode and the Glauber and soft modes are identical to the
new mode itself. Thus, after taking the overlap contributions into
account, the contributions from any additional modes vanish. A similar
argument can be made that no additional modes are required even if
power suppressed terms are considered.

\subsection{Relevance of Glauber gluons to other processes}%
\label{sec:GlaubersAndOtherProcesses}

In this Section we will assume that a generic one-loop diagram has a
pinch singularity in the Glauber region and derive some necessary
characteristics of the diagram from this and a few other
assumptions. The following analysis is closely related to the
preceding Section. Previously we started by studying a loop diagram with
given external leg momenta and found that the Glauber region is
pinched. Here we are going the opposite way: Let us start with the
assumptions that
\begin{enumerate}
\item the loop momentum $l \sim (\ll,\ll,\l)$ leads to a pinch singularity,
\item external momenta $k_i$ are on the mass-shell,
\item a momentum component can be at most $O(1)$, no inverse powers of $\l$.
\end{enumerate}

Our choice will be to perform the $l^+$ integration by
contours. Consider the propagator $(l+k)^{-2}$, which leads to a
simple pole of the form $(l^+-z)^{-1}$ with
\begin{equation}
z = \frac{(l_\perp+k_\perp)^2 - i 0}{l^-+k^-} - k^+\,.
\end{equation}
We will use the following ansatz for the scaling of $k$, namely $k^\pm
\sim O(\l^{n_\pm})$, $k_\perp \sim O(\l^{n_\perp})$. The on-shell
condition means that $n_+ + n_- = 2 n_\perp$. Since $l^+ \sim O(\ll)$,
it follows that $z$ must also scale like $O(\ll)$, and one of the
following conditions $(a)$ or $(b)$ must be satisfied:
\begin{eqnarray}
(a) && 2\, \hbox{min}(1,n_\perp) - \hbox{min}(2,n_-) = 2 \qquad \hbox{and} \qquad n_+ \ge 2\,, \nonumber \\
(b) && 2\, \hbox{min}(1,n_\perp) - \hbox{min}(2,n_-) > 2 \qquad \hbox{and} \qquad n_+ = 2\,. \nonumber
\end{eqnarray}
It is easy to show that condition $(b)$ can never be satisfied with
the assumption that $n_\pm, n_\perp \ge 0$, and that the only scaling
of $k$ consistent with $(a)$ is $k\sim (\ll,1,\l)$,
i.e.~collinear. Strictly speaking we get $n_- = 0$ and $n_\perp\ge 1$,
e.g.~$(\l^4,1,\ll)$ would also be allowed. The momentum with largest
invariant mass, however, (the least restrictive condition) is the
standard collinear one. External legs with even smaller invariant
masses will pinch the loop integral also in a Glauber-type
configuration, for example $l\sim (\l^4,\l^4,\l^2)$, which is readily
rescaled by $\lambda' = \ll$ to yield the same result.

\FIGURE{
\epsfig{file=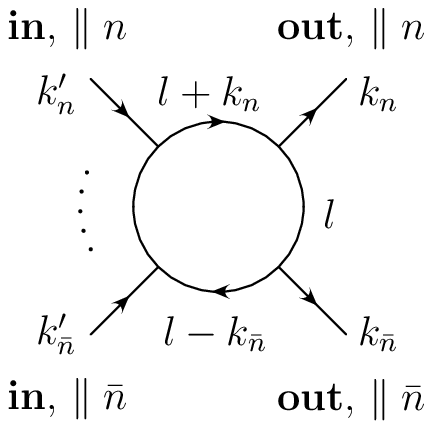, width=3.5cm}
\caption{\label{fig:anyLoop}A one-loop diagram with loop momentum $l$.}
%
}

We have therefore found that attaching an incoming collinear leg to
the loop is consistent with a Glauber loop momentum, although
certainly not sufficient. In particular, note that $k^-$ dominates
over $l^-$. In order to obtain a pinch singularity, we will hence need
another leg with collinear scaling, but with outgoing
momentum\footnote{The next propagator will then be $(l+K)^{-2}$ with
  $K = k-k'$. Since both $k$ and $k'$ have collinear scaling, so does
  $K$.}. Next, we note that it was a choice to perform the $l^+$
integration by contour, which pins down the $l^+$ scaling but leaves
$l^-$ open. We therefore repeat the above argument starting with the
$l^-$ component by contours and find that we need to attach two more
legs to the loop diagram, one in- and one outgoing with anti-collinear
momentum scaling.

It follows that any process involving diagrams of the form depicted in
Figure~\ref{fig:anyLoop} will be sensitive to Glauber
interactions. The characteristics are that there are at least four the
external legs, with two of them collinear and two of them
anticollinear. For a pinch singularity, each pair must consist of an
incoming and an outgoing momentum. The simplest such case is forward
scattering of a collinear with an anti-collinear field, where no other
external legs are attached to the diagram. Multiple exchanges of
Glauber gluons will give rise to an effective potential between the
collinear and anti-collinear fields when integrating Glauber gluons
out of the theory. This is similar to the exchange of potential gluons
in non-relativistic QCD, and can be found elsewhere in the literature
\cite{Pineda:1998kn, Luke:1999kz}. In our case of Drell-Yan far from machine
threshold there is one extra leg in the diagram from which the lepton
pair will emerge. From the point of view of Glauber gluon relevance
this case is identical to Higgs production, or indeed any New Physics
particle production in hadronic collisions away from machine
threshold.

\section{Conclusions and Outlook}\label{sec:concl}

Using an explicit matching calculation we have investigated whether
Glauber gluons are required as a degree of freedom in SCET. Focusing
on the operator $O_2$, whose Wilson coefficient is well known, we
showed that Glauber gluons are required if the matching is performed
with final states that contain both initial and final collinear
particles in the same collinear direction.

Even though we did not directly consider the Drell-Yan process, our
choice of external states was such that all contributions of the
Drell-Yan process, namely spectator-active and spectator-spectator
interactions in addition to active-active were present. This allows us
to make parallels between our conclusion about the consistency of
matching and the correct modes for the Drell-Yan amplitude. Our
conclusion is that for the exclusive Drell-Yan amplitude the correct
effective theory would require Glauber modes.  Note that we have not
discussed under what circumstances the contribution of Glauber gluons
cancel when squaring the amplitude.

For our analysis it was important to avoid double counting between the
modes by performing zero-bin subtractions \cite{Manohar:2006nz} from
the collinear and Glauber modes. It is worthwhile to emphasize the
spectator-active case for $\text{SCET}_G$, since in this topology
there is an interesting example of non-vanishing zero-bin subtraction
from the naive collinear mode $(I_4^{(n)c'})_{0g}\ne 0$. It shows that
taking these subtraction serious renders the effective theory robust
toward the introduction of other modes and does not invalidate
previous calculations.

From the scaling of the Glauber mode it is apparent that it cannot be
made exactly on-shell. However in the case of the spectator-spectator
topology there is a pinch singularity in the Glauber region. Thus we
find an apparent contradiction to the Coleman-Norton theorem
\cite{Coleman:1965xm}, which states that all pinched surfaces arise
from on-shell degrees of freedom. This contradiction is resolved if
one notices that Landau Equations used in the Coleman-Norton theorem
are conditions for appearance of a true singularity, which correspond
to the case when the loop integral is infinite, i.e when power
counting parameter of the effective theory vanishes: $\lambda=0$. In
this limit both the Glauber and soft momenta are identical to each
other $l^{\mu}_\text{g}=l^{\mu}_\text{s}=0$. Of course in this limit
both modes become on-shell, so the contradiction with Coleman-Norton
theorem goes away.

It would be interesting to reformulate the Landau equations taking
into account the proper power counting, i.e. instead of writing down a
condition of having a true pinch singularity, find a condition for
pinched poles to occur at distance of say order $\lambda^2$ from each
other. It should be possible to use such an analysis to discover both
Glauber and soft pinches in the spectator-spectator diagram directly
from such relaxed Landau Equations.  Understanding this question in
details might potentially be of practical importance since it can
lead to a nice recipe for arbitrary process on how to read off the
correct long distance modes (valid to all orders in perturbation
theory with a possibility to include power corrections). This,
however, is beyond the scope of the present paper and will be studied
elsewhere.

Another important step is to study the effect of Glauber gluons in
SCET on physical observables, and not just amplitudes as done in this
paper. The expectation here would be to understand the cancellation of Glauber gluons
in the inclusive in the final hadronic state Drell-Yan cross-section, which was proved in
 full QCD in Refs. \cite{Collins:1982wa,Bodwin:1984hc,Collins:1985ue}. The main challenge in doing so is that the Glauber mode scaling
is such that the corresponding particle is always off-shell, such that
it can not be naively included in the SCET Lagrangian. A better way to
proceed might be to interpret this mode as an effective potential,
giving rise to forward scattering of two collinear particles in
different directions~\cite{StewartTalk}.

\acknowledgments We would like thank Thomas Becher for helpful
discussions, and Christopher Lee, Zoltan Ligeti, George Sterman and
Iain Stewart for comments on the manuscript. This work was supported
by the Director, Office of Science, Offices of High Energy and Nuclear
Physics of the U.S.  Department of Energy under the Contract
DE-AC02-05CH11231. G.O. was additionally supported under the Contract DE-AC52-06NA25396 and in part by the LDRD program
at LANL.  B.L. would like to thank the University of Siegen for their hospitality during the final stages of this project. 

\bibliography{bibliography}

\providecommand{\href}[2]{#2}\begingroup\raggedright\begin{thebibliography}{10}

\bibitem{Amati:1978by}
D.~Amati, R.~Petronzio, and G.~Veneziano {\em Nucl. Phys.} {\bf B146} (1978)
  29--49.

\bibitem{Libby:1978qf}
S.~B. Libby and G.~F. Sterman {\em Phys. Rev.} {\bf D18} (1978) 3252.

\bibitem{Mueller:1978xu}
A.~H. Mueller {\em Phys. Rev.} {\bf D18} (1978) 3705.

\bibitem{Gupta:1979xj}
S.~Gupta and A.~H. Mueller {\em Phys. Rev.} {\bf D20} (1979) 118.

\bibitem{Ellis:1978ty}
R.~K. Ellis, H.~Georgi, M.~Machacek, H.~D. Politzer, and G.~G. Ross {\em Nucl.
  Phys.} {\bf B152} (1979) 285.

\bibitem{Sterman:1978bi}
G.~F. Sterman {\em Phys. Rev.} {\bf D17} (1978) 2773.

\bibitem{Sterman:1978bj}
G.~F. Sterman {\em Phys. Rev.} {\bf D17} (1978) 2789.

\bibitem{Collins:1981ta}
J.~C. Collins and G.~F. Sterman {\em Nucl. Phys.} {\bf B185} (1981) 172.

\bibitem{Collins:1981uk}
J.~C. Collins and D.~E. Soper {\em Nucl. Phys.} {\bf B193} (1981) 381.

\bibitem{Collins:1987pm}
J.~C. Collins and D.~E. Soper {\em Ann. Rev. Nucl. Part. Sci.} {\bf 37} (1987)
  383--409.

\bibitem{Collins:1989gx}
J.~C. Collins, D.~E. Soper, and G.~F. Sterman {\em Adv. Ser. Direct. High
  Energy Phys.} {\bf 5} (1988) 1--91,
  [\href{http://arxiv.org/abs/hep-ph/0409313}{{\tt hep-ph/0409313}}].

\bibitem{Landau:1959fi}
L.~D. Landau {\em Nucl. Phys.} {\bf 13} (1959) 181--192.

\bibitem{Coleman:1965xm}
S.~Coleman and R.~E. Norton {\em Nuovo Cim.} {\bf 38} (1965) 438--442.

\bibitem{Bodwin:1981fv}
G.~T. Bodwin, S.~J. Brodsky, and G.~P. Lepage {\em Phys. Rev. Lett.} {\bf 47}
  (1981) 1799.

\bibitem{Collins:1981tt}
J.~C. Collins, D.~E. Soper, and G.~F. Sterman {\em Phys. Lett.} {\bf B109}
  (1982) 388.

\bibitem{Collins:1982wa}
J.~C. Collins, D.~E. Soper, and G.~Sterman {\em Nucl. Phys.} {\bf B223} (1983)
  381.

\bibitem{Bodwin:1984hc}
G.~T. Bodwin {\em Phys. Rev.} {\bf D31} (1985) 2616 [Erratum: Phys. Rev. D 34
  (Dec, 1986), 3932].

\bibitem{Collins:1985ue}
J.~C. Collins, D.~E. Soper, and G.~F. Sterman {\em Nucl. Phys.} {\bf B261}
  (1985) 104.

\bibitem{Aybat:2008ct}
S.~M. Aybat and G.~F. Sterman {\em Phys. Lett.} {\bf B671} (2009) 46--50,
  [\href{http://arxiv.org/abs/0811.0246}{{\tt arXiv:0811.0246}}].

\bibitem{Bauer:2002nz}
C.~W. Bauer, S.~Fleming, D.~Pirjol, I.~Z. Rothstein, and I.~W. Stewart {\em
  Phys. Rev.} {\bf D66} (2002) 014017,
  [\href{http://arxiv.org/abs/hep-ph/0202088}{{\tt hep-ph/0202088}}].

\bibitem{Bauer:2002ie}
C.~W. Bauer, A.~V. Manohar, and M.~B. Wise {\em Phys. Rev. Lett.} {\bf 91}
  (2003) 122001, [\href{http://arxiv.org/abs/hep-ph/0212255}{{\tt
  hep-ph/0212255}}].

\bibitem{Bauer:2003di}
C.~W. Bauer, C.~Lee, A.~V. Manohar, and M.~B. Wise {\em Phys. Rev.} {\bf D70}
  (2004) 034014, [\href{http://arxiv.org/abs/hep-ph/0309278}{{\tt
  hep-ph/0309278}}].

\bibitem{Manohar:2003vb}
A.~V. Manohar {\em Phys. Rev.} {\bf D68} (2003) 114019,
  [\href{http://arxiv.org/abs/hep-ph/0309176}{{\tt hep-ph/0309176}}].

\bibitem{Gao:2005iu}
Y.~Gao, C.~S. Li, and J.~J. Liu {\em Phys. Rev.} {\bf D72} (2005) 114020,
  [\href{http://arxiv.org/abs/hep-ph/0501229}{{\tt hep-ph/0501229}}].

\bibitem{Idilbi:2005er}
A.~Idilbi, X.-d. Ji, and F.~Yuan {\em Phys. Lett.} {\bf B625} (2005) 253--263,
  [\href{http://arxiv.org/abs/hep-ph/0507196}{{\tt hep-ph/0507196}}].

\bibitem{Idilbi:2005ky}
A.~Idilbi and X.-d. Ji {\em Phys. Rev.} {\bf D72} (2005) 054016,
  [\href{http://arxiv.org/abs/hep-ph/0501006}{{\tt hep-ph/0501006}}].

\bibitem{Chay:2005rz}
J.~Chay and C.~Kim {\em Phys. Rev.} {\bf D75} (2007) 016003,
  [\href{http://arxiv.org/abs/hep-ph/0511066}{{\tt hep-ph/0511066}}].

\bibitem{Manohar:2005az}
A.~V. Manohar {\em Phys. Lett.} {\bf B633} (2006) 729--733,
  [\href{http://arxiv.org/abs/hep-ph/0512173}{{\tt hep-ph/0512173}}].

\bibitem{Becher:2006mr}
T.~Becher, M.~Neubert, and B.~D. Pecjak {\em JHEP} {\bf 01} (2007) 076,
  [\href{http://arxiv.org/abs/hep-ph/0607228}{{\tt hep-ph/0607228}}].

\bibitem{Chen:2006vd}
P.-y. Chen, A.~Idilbi, and X.-d. Ji {\em Nucl. Phys.} {\bf B763} (2007)
  183--197, [\href{http://arxiv.org/abs/hep-ph/0607003}{{\tt hep-ph/0607003}}].

\bibitem{Lee:2006nr}
C.~Lee and G.~F. Sterman {\em Phys. Rev.} {\bf D75} (2007) 014022,
  [\href{http://arxiv.org/abs/hep-ph/0611061}{{\tt hep-ph/0611061}}].

\bibitem{Fleming:2007qr}
S.~Fleming, A.~H. Hoang, S.~Mantry, and I.~W. Stewart {\em Phys. Rev.} {\bf
  D77} (2008) 074010, [\href{http://arxiv.org/abs/hep-ph/0703207}{{\tt
  hep-ph/0703207}}].

\bibitem{Fleming:2007xt}
S.~Fleming, A.~H. Hoang, S.~Mantry, and I.~W. Stewart {\em Phys. Rev.} {\bf
  D77} (2008) 114003, [\href{http://arxiv.org/abs/0711.2079}{{\tt
  arXiv:0711.2079}}].

\bibitem{Becher:2007ty}
T.~Becher, M.~Neubert, and G.~Xu {\em JHEP} {\bf 07} (2008) 030,
  [\href{http://arxiv.org/abs/0710.0680}{{\tt arXiv:0710.0680}}].

\bibitem{Schwartz:2007ib}
M.~D. Schwartz {\em Phys. Rev.} {\bf D77} (2008) 014026,
  [\href{http://arxiv.org/abs/0709.2709}{{\tt arXiv:0709.2709}}].

\bibitem{Bauer:2008dt}
C.~W. Bauer, S.~P. Fleming, C.~Lee, and G.~F. Sterman {\em Phys. Rev.} {\bf
  D78} (2008) 034027, [\href{http://arxiv.org/abs/0801.4569}{{\tt
  arXiv:0801.4569}}].

\bibitem{Bauer:2008jx}
C.~W. Bauer, A.~Hornig, and F.~J. Tackmann {\em Phys. Rev.} {\bf D79} (2009)
  114013, [\href{http://arxiv.org/abs/0808.2191}{{\tt arXiv:0808.2191}}].

\bibitem{Bauer:2010vu}
C.~W. Bauer, N.~D. Dunn, and A.~Hornig
  \href{http://arxiv.org/abs/1002.1307}{{\tt arXiv:1002.1307}}.

\bibitem{Bauer:2000ew}
C.~W. Bauer, S.~Fleming, and M.~E. Luke {\em Phys. Rev.} {\bf D63} (2000)
  014006, [\href{http://arxiv.org/abs/hep-ph/0005275}{{\tt hep-ph/0005275}}].

\bibitem{Bauer:2000yr}
C.~W. Bauer, S.~Fleming, D.~Pirjol, and I.~W. Stewart {\em Phys. Rev.} {\bf
  D63} (2001) 114020, [\href{http://arxiv.org/abs/hep-ph/0011336}{{\tt
  hep-ph/0011336}}].

\bibitem{Bauer:2001ct}
C.~W. Bauer and I.~W. Stewart {\em Phys. Lett.} {\bf B516} (2001) 134--142,
  [\href{http://arxiv.org/abs/hep-ph/0107001}{{\tt hep-ph/0107001}}].

\bibitem{Bauer:2001yt}
C.~W. Bauer, D.~Pirjol, and I.~W. Stewart {\em Phys. Rev.} {\bf D65} (2002)
  054022, [\href{http://arxiv.org/abs/hep-ph/0109045}{{\tt hep-ph/0109045}}].

\bibitem{Liu:2008cc}
F.~Liu and J.~P. Ma \href{http://arxiv.org/abs/0802.2973}{{\tt
  arXiv:0802.2973}}.

\bibitem{Idilbi:2008vm}
A.~Idilbi and A.~Majumder {\em Phys. Rev.} {\bf D80} (2009) 054022,
  [\href{http://arxiv.org/abs/0808.1087}{{\tt arXiv:0808.1087}}].

\bibitem{DEramo:2010ak}
F.~D'Eramo, H.~Liu, and K.~Rajagopal \href{http://arxiv.org/abs/1006.1367}{{\tt
  arXiv:1006.1367}}.

\bibitem{StewartTalk}
I.~Stewart and I.~Rothstein (work in progress). Talk ``Glauber Gluons in SCET"
  presented by I. Stewart at SCET2010 in Ringberg, Germany, 2010.

\bibitem{Bauer:2008qu}
C.~W. Bauer, O.~Cata, and G.~Ovanesyan
  \href{http://arxiv.org/abs/0809.1099}{{\tt arXiv:0809.1099}}.

\bibitem{Manohar:2006nz}
A.~V. Manohar and I.~W. Stewart {\em Phys. Rev.} {\bf D76} (2007) 074002,
  [\href{http://arxiv.org/abs/hep-ph/0605001}{{\tt hep-ph/0605001}}].

\bibitem{Pineda:1998kn}
A.~Pineda and J.~Soto {\em Phys. Rev.} {\bf D59} (1999) 016005,
  [\href{http://arxiv.org/abs/hep-ph/9805424}{{\tt hep-ph/9805424}}].

\bibitem{Luke:1999kz}
M.~E. Luke, A.~V. Manohar, and I.~Z. Rothstein {\em Phys. Rev.} {\bf D61}
  (2000) 074025, [\href{http://arxiv.org/abs/hep-ph/9910209}{{\tt
  hep-ph/9910209}}].

\end{thebibliography}\endgroup

\end{document}